\documentclass{aa}

\usepackage{graphicx}
\usepackage{txfonts}
\usepackage{epsfig}
\usepackage{color}

\newcommand{\Te}{$T_\textrm{e}$} 
\newcommand{\Fx}{$F_\textrm{x}$} 

\usepackage[T1]{fontenc}
\usepackage{ae,aecompl}

\usepackage{mathptmx}
\usepackage{txfonts}

\begin{document}


\title{Hard X-ray variability of V404 Cyg during the 2015 outburst\thanks{Based on observations with INTEGRAL, an ESA project with instruments and science data centre funded by ESA member states (especially the PI countries: Denmark, France, Germany, Italy, Switzerland, Spain) and with the participation of Russia and the USA.}}

\author{C. S\'anchez-Fern\'andez\inst{1}\and
J.~J.~E. Kajava\inst{1,2,3}\and
S.~E. Motta\inst{4}\and
E. Kuulkers\inst{1,5}
}

\institute{European Space Astronomy Centre (ESA/ESAC), Science Operations Department, E-28691, Villanueva de la Ca\~{n}ada, Madrid, Spain \email{Celia.Sanchez@sciops.esa.int} \and 
Finnish Centre for Astronomy with ESO (FINCA), University of Turku, V\"{a}is\"{a}l\"{a}ntie 20, FIN-21500 Piikki\"{o}, Finland
\and Tuorla Observatory, Department of Physics and Astronomy, University of Turku, V\"{a}is\"{a}l\"{a}ntie 20, FIN-21500 Piikki\"{o}, Finland \and
University of Oxford, Department of Physics, Astrophysics, Denys Wilkinson Building, Keble Road, Oxford OX1 3RH, UK
\and ESA/ESTEC, Keplerlaan 1, 2201 AZ Noordwijk, The Netherlands\\}

   \date{Received August, 2016; accepted ..., ...}

  \abstract
{}
{Hard X-ray spectra of black hole binaries (BHB) are produced by Comptonization of soft seed photons by hot electrons near the black hole. The slope of the resulting energy spectra is governed by two main parameters: the electron temperature (\Te) and the optical depth ($\tau$) of the emitting plasma. Given the extreme brightness of V404~Cyg during the 2015 outburst, we aim to constrain the source spectral properties using an unprecedented time resolution in hard X-rays, and to monitor the evolution of \Te\ and $\tau$ over the outburst.}
{We have extracted and analyzed 602 X-ray spectra of V404 Cyg obtained by the IBIS/ISGRI instrument on-board \textit{INTEGRAL} during the 2015 June outburst, using effective integration times  ranging between 8 and 176000 seconds. We have fit the resulting spectra  in the 20--200\,keV energy range.}
{We find that while the  light curve and soft X-ray spectra of V404 Cyg are remarkably different from those of other BHBs, the spectral evolution of V404~Cyg in hard X-rays and the relations between the spectral parameters are consistent with those observed in other BHBs. We identify a \textit{hard branch} where the \Te\ is  anti-correlated with the hard X-ray flux, and a \textit{soft flaring branch} where the relation reverses. In addition, we find that during long  X-ray plateaus detected at intermediate fluxes, the thermal Comptonization models fail to describe the spectra. However, the statistics improve if we allow $N_\textrm{H}$ to vary freely in the fits to these spectra.} 
{We conclude that the \textit{hard branch} in V404 Cyg is analogous to the canonical hard state of BHBs. V404~Cyg never seems to enter the canonical soft state, although the soft flaring branch bears resemblance to the BHB intermediate state and ultra-luminous state. The X-ray plateaus are likely 
the result of absorption by a Compton-thick outflow ($N_\textrm{H} \gtrsim 10^{24}\,\textrm{cm}^{-2}$) which reduces the observed flux by a factor of about 10. Variable covering of the central source by this Compton-thick material may be the reason for the complicated light curve variability, rather than intrinsic source variability.}
\maketitle

\section{Introduction}

\noindent
Black hole (BH) binary systems (BHBs) can go through various spectral states which are thought to be caused  by changes in the accretion geometry and accretion rates close to the BH, although the actual details still remain debated \citep{RMcC06,BM16}.
The two most common states are the \textit{hard} and the \textit{soft} states (see, e.g., \citealt{DGK07,PV14} for review).
In the \textit{hard state}, the spectrum can be described by a power-law with a variable cut-off energy around 60-150\,keV, which is thought to result from  Comptonization of soft seed photons by a population of hot electrons located in an optically thin  region  close to the BH \citep{SLE76,NY95}. The high energy cut-off suggests a thermal distribution of electrons, with temperatures in the range  \mbox{30--100\,~keV} \citep{ST79,GZP99}. Occasionally, a hard excess has been observed above 100~keV suggesting  the presence  of non-thermal electrons as well (see e.g., \citealt{MZB02, WZG02, JJM07, DBM10}), either in the corona/hot flow or in the base of the jet (e.g., \citealt{ZLS12}). 
In the \textit{soft state}, thermal emission peaking at $\sim$\,1\,keV, from a cool, optically thick, geometrically thin accretion disk,  dominates the spectrum \citep{SS73,EMcCN97}. A weak,  hard X-ray tail extending up to the MeV range, is also detected \citep{ZMC16}. This tail is thought to originate from Comptonized emission by non-thermal electrons in discrete flares on top of the accretion disc \citep{MZB02}.
During transitions between the hard and soft states, BHBs  pass through additional  \textit{intermediate} states, which show characteristic features of both (see e.g. \citealt{EOA94,MPJ06,BM16}).
On rare occasions, some systems may pass also through the so-called \textit{ultra-luminous state} \citep{MHM12}, also called {\it very high state} or {\it anomalous} state,
an intermediate state characterized by both a strong  thermal component and a very strong and steep hard X-ray tail \citep{DGK07}.\\
\indent One of the current observational challenges in this context is to determine the  electron temperature, \Te, and optical depth, $\tau$, of the Comptonizing medium. These together determine the spectral slope of the  Comptonized spectrum (e.g. \citealt{B99ASP}).
Because the cut-off energies are found around 100\,keV, where usually the instrumental response is low, 
observations sensitive enough to constrain these parameters have only been available for a few sources and typically require long exposures.
Observations in the \textit{hard state} of GRO J0422+32 \citep{ENC98},  GX~339--4  \citep{WZG02,MBH09}, XTE J1550--564 \citep{RCT03}, Cyg X-1 \citep{dSMB13} and Swift~J1753.5--0127 \citep{KVT16} show an anti-correlation between the electron temperature, \Te\ (or high energy cut-off) and the X-ray flux, accompanied by 
a correlation between the plasma optical depth, $\tau$, and the X-ray flux \citep{WZG02}.  
These relations reverse during the \textit{hard} to \textit{soft} state transitions. Observations of these transitions in  Cyg~X-1 \citep{PJL96,dSMB13}, GRO J1719--24 \citep{ENC98}, GRO~J1655--40 \citep{JKS08} and GX~339--4 \citep{MBH09} show an increasing \Te\ with increasing  flux, while the optical depth $\tau$ decreases \citep{JKS08,dSMB13}.  The cut-off is significantly present during the hard  and intermediate states,  and it disappears when the source reaches the soft state.

The extremely bright outburst of V404 Cyg in June 2015 provides a unique data set to perform high time resolved spectroscopy in high energies and study in detail the evolution of the parameters describing the Comptonizing plasma. We present here the results of spectral analysis of IBIS/ISGRI data in the 20--200\,keV energy range  over the period 18--28 June 2015, when the source was brightest.

V404~Cyg is a  transient  Low-Mass X-ray Binary (LMXB) consisting of a $9.0^{+0.2}_{-0.6}\,\textrm{M}_{\odot}$ BH
 accreting mass from a  K3~III companion \citep{KFR10} in a 6.5\,d orbit \citep{CC92}. It is located at a distance $d\!=\!2.39 \pm 0.14$~kpc \citep{MJJD09}.
V404~Cyg was first detected in optical wavelengths during two outbursts in 1938 and 1956 \citep{R89} and later in X-rays during a third outburst in 1989 \citep{Mk89, Mr89}. The 1989 outburst was
characterized by 
extreme flaring activity, several flux levels above the Crab  \citep{T89,OvKV96}. 
After $\sim$\,26 years in quiescence, the onset of a new outburst was detected by {\it Swift}/BAT, MAXI and \textit{Fermi}/GBM on 15 June 2015 \citep{BS15,NMM15,Y15}. This outburst, which triggered the most intensive multi-wavelength observing campaign performed so far on a transient  BHB, lasted until early-August 2015 \citep{SBA15}.  
During the first ten days, the source exhibited violent flaring activity on time scales of sub-seconds to hours in all wavelengths:
$\gamma$-rays \citep{LCD16}; 
X-rays (\citealt{RCBA15,RJB15,JWHH16}; \citeauthor{W16} et al. \citeyear{W16}); optical \citep{Getal16,KIK16,MDCMS16}; infrared \citep{EDG16}; millimeter/sub-millimeter and radio \citep{TSY15}. In some major flares, V404 Cyg reached fluxes around 50 and 40 Crab in soft and hard X-rays, respectively \citep{SDSDA15,RCBA15}. The peak of the outburst was reached on June 26th, and the flux dropped immediately afterwards \citep{FBB15,WHF15} slowly fading to quiescence over the subsequent weeks \citep{SBA15}.

\begin{figure*}[t]
\begin{tabular}{lll}
\includegraphics{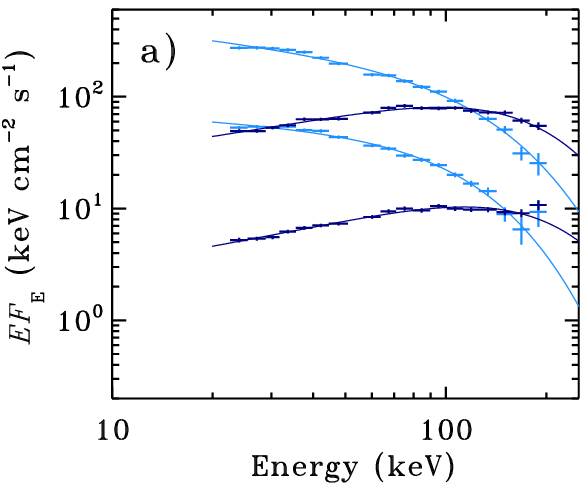} & \hspace{-0.2cm}
\includegraphics{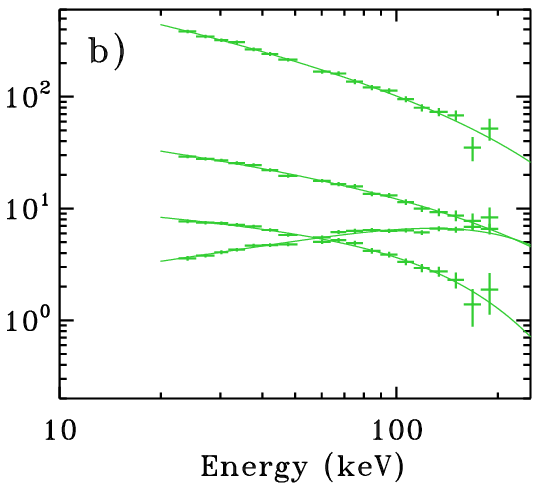} & \hspace{-0.2cm}
\includegraphics{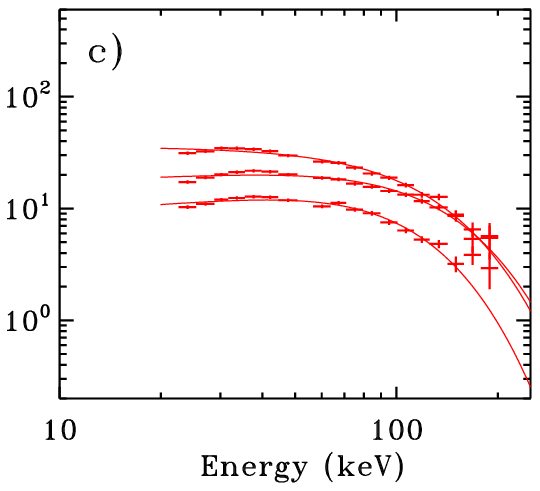}
\end{tabular}
\caption{Example of the spectra analyzed in this work. {\bf Panel a)} Comptonized spectra displaying a cutoff at high energies within the IBIS/ISGRI energy range. We classify these spectra in two groups: {\sl hard} ($\Gamma$\,$\le$\,1.7; dark blue) and {\sl soft} ($\Gamma$\,$>$\,1.7; light blue). The hardness selection is based on the  $\Gamma$ values derived using the {\sc nthcomp} model. From top to bottom, the effective integration times for these spectra are 36, 170 seconds ($\Gamma$\,$>$\,1.7 spectra) and 91, 1285 seconds ($\Gamma$\,$\le$\,1.7 spectra).  {\bf Panel b)} Comptonized spectra for which \Te\ cannot be constrained by our data. We fixed the electron temperature to the value \Te\ =\,999\,keV in these spectral fits. Note the range in fluxes and hardness presented by these spectra. The effective integration times for these spectra are 8, 403, 1730 and 2251 seconds (top to bottom). {\bf Panel c)} spectra for which a Comptonized model was not statistically favored by our model selection criteria ($p$--value $<$\,0.05). These spectra are predominantly found during the X-ray plateaus detected at fluxes \Fx\ $\sim\!5\times\!10^{-8}$\,erg\,cm$^{-2}$\,s$^{-1}$, as explained in the text.  The effective integration times for these spectra are 297, 579 and 912 seconds (top to bottom).}
\label{fig:spe.example}
\end{figure*}

\begin{figure*}
\includegraphics[width=22.5cm,angle=90]{figures/Overall_lcr_400mil.eps}
\vspace{-0.35cm}
\caption{Evolution of the flux and spectral parameters of V404~Cyg  during the June 2015 flaring episodes. \textbf{Panel a)} source flux (20--200 keV) in units of 10$^{-8}$\,erg\,cm$^{-2}$\,s$^{-1}$. \textbf{Panel b, c, d)} {\sc nthcomp} fitting parameters (power-law index, $\Gamma$, electron temperature, \Te\  and $\chi^{2}_{\nu}$). \textbf{Panel e, f, g) } {\sc compps} fitting parameters (optical depth, $\tau$,   electron temperature, \Te\,  and $\chi^{2}_{\nu}$). 
Green, blue and red symbols are used to highlight  the  best-fitting model, according to our model selection criteria (Sect. \ref{sec:spectral.modelling}). Blue: Comptonization models with  constrained \Te\ (Fig.~\ref{fig:spe.example}a) further divided into hard  ($\Gamma\,\le\,1.7$; dark blue) and soft spectra  ($\Gamma$\,$>$\,1.7; light blue). Green: Comptonization models with unconstrained \Te\ (Fig.~\ref{fig:spe.example}b). Red: Comptonization models with $p$--value $<$0.05 fits.  (Fig.~\ref{fig:spe.example}c). Note that $\chi^{2}_{\nu}$ $\gtrsim$ 2 values are only obtained during the X-ray plateaus observed at \Fx $\sim$ 5$\times$\,10$^{-8}$\,erg\,cm$^{-2}$\,s$^{-1}$.}
\label{fig:overall_lcr}
\end{figure*}

\section{Observations and data analysis}

V404 Cyg was observed by {\sl INTEGRAL}, the INTErnational Gamma-Ray Astrophysics Laboratory \citep{W03} in a series of Target of Opportunity observations scheduled between 17 June, 2015 and 13 July, 2015 (MJD 57556--57582; revolutions 1554--1563; \citealt{K15}).
We present here the analysis of the available IBIS/ISGRI data \citep{LLL03}, obtained during revolutions 1554--1558 (18--28 June, 2015; MJD 57191--51201), which cover the epoch of intense flaring activity and the beginning of the outburst decay. These observations provide data sensitive enough to study in detail the properties of the Comptonizing medium. 

\subsection{Data reduction}
The IBIS/ISGRI data reduction was performed using the Off-line Scientific Analysis software (OSA; \citealt{CWB03}) v10.2, using the latest calibration files. The data were processed following standard IBIS/ISGRI  reduction procedures.

The spectral extraction was performed using good time interval files (GTIs) of variable duration, defined to provide source spectra of comparable S/N regardless of the source flux. The GTI selection was based on the source light curves distributed by the INTEGRAL Science Data Center (ISDC; \citealt{K15}). The GTIs were defined sequentially, and their start/end times were selected such that during  each time interval  4$\times$10$^5$ counts were accumulated in the IBIS/ISGRI 25--60\, keV band. This GTI selection was found to be an optimal compromise between time resolution and the ability to constrain the \Te\ values, particularly for the softer spectra in the sample. Using this strategy, 
we extracted 602 spectra, with effective exposure times in the range  8 to 176000\,s.
We binned the IBIS/ISGRI response matrix in the energy range 20--500\,keV using 28 channels of variable logarithmic widths. To remove potential background contamination, and the contribution of additional  spectral components above 200\,keV, like  hard non-thermal tails \citep{RCBA15,RJB15} or  hard X-ray emission caused by positron annihilation \citep{SDG16}, we  restricted the spectral fits to the 20--200\,keV energy range. In our fits we have ignored the energy bin around 50\,keV due to calibration uncertainties and added 3~per~cent systematic errors to the spectral bins.
The IBIS/ISGRI X-ray spectra were fit using \textsc{xspec} v12.8.2 \citep{A96}, adopting the $\chi^2$~statistics. Errors provided below are quoted at the 1-$\sigma$~confidence level ($\Delta\chi^2=1$ for one parameter of interest).
 
\subsection{Spectral modelling}
\label{sec:spectral.modelling}

We show in Fig.~\ref{fig:spe.example} a sample of the IBIS/ISGRI spectra of V404~Cyg analysed in this work. In most occasions,  the spectra have a powerlaw-like shape, modified by a cut-off at high energies,  consistent with a thermal Comptonization spectrum (see Fig.~ \ref{fig:spe.example}a).
Therefore, we have used  Comptonization models to fit our data:
{\sc nthcomp} \citep{ZJM96, ZDS99}  and {\sc compps} \citep{PS96}. These models provide a description of the continuum produced by thermal Compton up-scattering of soft X-ray photons.
{\sc nthcomp} 
is parameterized by a power law index $\Gamma$, and an electron temperature \Te. The 
{\sc compps} parameters are the electron temperature \Te\ and optical depth $\tau$.
Although a thermal component was never detected during the June 2015 outburst (Motta et al. in preparation), in our fits
 we fixed the seed photon temperature $T_{\rm bb}$\,=\,0.1~keV, as fits by \cite{MKS16} to the source spectra over the 0.6--200\,keV energy range, using Comptonization models are consistent with this value.

In some cases, a high-energy cutoff is either weakly significant or not statistically required by the data (see figure \ref{fig:spe.example}b). These spectra can still be fit using Comptonization models, but fixing \Te\ to an arbitrary high value (\Te\,=\,999\,keV).
To account for both possibilities, we carried out two independent fitting runs per model. In the first fitting run, we fit every spectrum leaving \Te\ as a free parameter,
while in the second fitting run we fixed it to \Te\,=\,999\,keV. Then, the  Bayesian information criterion (BIC; \citealt{S78}) was independently applied to the results obtained for every
 spectrum, in order to select the best fit to the data.
We computed the BIC using the following approximation:  
BIC\,=\,$\chi^2 + k \ln(n)$,  
where $k$ is the number of 
parameters in the model, and $n$ is the number of channels in the spectral fits.
In a model selection process, the optimal  model is
identified by the minimum value of BIC. A lower BIC implies either fewer explanatory variables, a better fit, or both. 
\citet{KR95} set the strength of the evidence against the model with the higher BIC to be 
strong if $\Delta$BIC$>$6, which we adopted as the limit for model selection. This approach was applied to the {\sc nthcomp} and {\sc compps} fits. 
The results of this analysis are described in Sections 3.1--3.3, and displayed in figures  \ref{fig:overall_lcr}--8.
In these  figures the data are presented according to the following color convention:

- Blue points are used to highlight those fits where the $\Delta$BIC model selection  favored a Comptonization model with a constrained electron temperature \Te, further divided in  two groups: 'hard spectra' ($\Gamma$\,$\le$\,1.7; dark blue) and 'soft spectra'  ($\Gamma$\,$>$\,1.7; light blue). Some  of these spectra are shown in Fig.\,\ref{fig:spe.example}a.
The latter classification is based on the $\Gamma$ values derived from the {\sc nthcomp} fits, and then  applied to the {\sc compps} fits. 

- Green points correspond to those  spectra where  \Te\ could not be constrained by our fits (i.e. \Te\ fixed at 999\,keV). Some  of these spectra are shown in Fig.\, \ref{fig:spe.example}b.

- Additionally, we computed the corresponding $p$--value of the fit with respect to the data for every fit. We mark the spectra where $p <0.05$ with red symbols. Some  of these spectra are shown in Fig.\,\ref{fig:spe.example}c. 

To improve the fits to the latter group of spectra, we also explored the possibility that these were affected by heavy absorption. The results of these additional fits are presented in Sect. \ref{Var_abs}
 
\section{Results}
\label{sec:results}

\subsection{Parameter evolution}
\label{sect:params.evolution}

We present in Fig. \ref{fig:overall_lcr} the time evolution of the source flux computed in the 20--200 keV energy range, together with the evolution of the spectral parameters  derived using  {\sc nthcomp} ($\Gamma$,~\Te) and  {\sc compps}  ($\tau$,~\Te) and the corresponding $\chi^2_{\nu}$ values. 

Hereafter we will refer to the flux in the 20--200 keV energy range as~\Fx. 

\subsubsection{{\sc epoch 1}: Flaring activity}
\label{sect:epoch1}

During the period MJD 57191--57193 (Rev. 1554; {\sc epoch\,1} in Fig.~\ref{fig:overall_lcr}a) intense flaring activity was detected on timescales of minutes to hours. The flares reached  peak fluxes of \Fx$\sim$\,55\,$\times$\,10$^{-8}$\,erg\,cm$^{-2}$\,s$^{-1}$, while between flares we measure  fluxes below  \Fx$\sim\!2\!\times\!10^{-8}$\,erg\,cm$^{-2}$\,s$^{-1}$. Over this period, the source spectrum was hard ($\Gamma\,\le\,1.7$), and only softened when \Fx\ increased above  $\gtrsim$\,25\,$\times$\,10$^{-8}$\,erg\,cm$^{-2}$\,s$^{-1}$ (i.e. during the peaks of the flares).
\Te\ is well constrained during the X-ray flares, with values in the range  30--100\,keV ({\sc nthcomp}) or 30--70\,keV ({\sc compps}). Between flares \Te\ cannot be constrained in our spectral fits, and the X-ray spectrum is consistent with a hard   power-law ($\Gamma\,\le\,1.7$) with no cut-off (see Fig. \ref{fig:spe.example}b). Similar results were obtained by \cite{NFB15}, who analyzed this data set using a different time resolution, and \cite{RJB15}, who analyzed contemporaneous \textit{INTEGRAL}/SPI data.

In  {\sc epoch\,1} $\tau$ varied between 2 and 5. It displayed higher values  ($\tau$\,$\gtrsim$\,3.5) when the spectrum was hard ($\Gamma\,\le\,1.7$), and decreased  ($\tau$\, $\lesssim$\,3.5) as the spectrum softened ($\Gamma\,>\,1.7$; see Fig. \ref{fig:overall_lcr}d).

\subsubsection{{\sc epoch 2}: Spectral softening, state transitions, and  X-ray plateaus}
\label{sect:epoch2}
During the intervals MJD 57193.5--57195.5 and MJD 57196.7-- 57197.4 (Rev. 1555--1556; {\sc epoch 2} in Fig. \ref{fig:overall_lcr}a) the flaring activity persisted. Peak fluxes of $F_{\rm p}$\,$\sim$\,60--80\,$\times$\,10$^{-8}$\,erg\,cm$^{-2}$\,s$^{-1}$ were measured. Between flares, we find again fluxes $F_\textrm{x}\!\lesssim\!2\!\times\!10^{-8}$\,erg\,cm$^{-2}$\,s$^{-1}$.
In general, in {\sc epoch\,2} V404~Cyg  displayed softer spectra than during {\sc epoch\,1} even at the lowest count rates, when spectra with $\Gamma$\,$\sim$\,2.3 without a cut-off were frequently observed.  Only one of the X-ray flares (detected around MJD 57195.15) displayed a hard X-ray spectrum ($\Gamma\,\le$\,1.7).   Several  transitions between Comptonized hard ($\Gamma\,\le$\,1.7) and soft ($\Gamma$\,$\sim$\,3) spectra   with unconstrained \Te\ occurred. 

In {\sc epoch 2}, there is more scatter in the measured \Te, $\tau$ and $\Gamma$ parameters than in \textsc{epoch 1}. Also, the relation between \Fx\ and $\Gamma$ is complex: while during some flares $\Gamma$ was roughly constant (e.g. flare on MJD 57194.10, Fig. \ref{fig:zoom_plateau}), in other flares the spectrum hardened (e.g. flare on MJD 57195.5)
or softened during the entire flare (e.g. flare on MJD 57194.3, Fig. \ref{fig:zoom_plateau}). 
During the flares, \Te\ displayed values in the range  30--120\,keV ({\sc nthcomp}) or 20--150\,keV ({\sc compps}). 
Very soft spectra ($\Gamma\sim\!3$) with unconstrained \Te\ were detected by the end of \textsc{epoch 2}, when the flux  dropped below $\sim\!2\times 10^{-8}$\,erg\,cm$^{-2}$\,s$^{-1}$.

One additional feature over this period is the detection of X-ray plateaus (i.e. non-varying flux periods) at intermediate fluxes (\Fx\! $\sim\!5\!\times\!10^{-8}$\,erg\,cm$^{-2}$\,s$^{-1}$). They last several hours and happen in  between successive X-ray flares (red points in Fig. \ref{fig:overall_lcr}; see also \citealt{RCBA15}). For a closer view of one of these  plateaus see Fig.~\ref{fig:zoom_plateau}.
We consistently  obtain $p$--values\,$<$\,0.05 and $\chi_{\nu}^2>1.8$ when modeling the plateau spectra using Comptonization models ({\sc nthcomp} or {\sc compps}), which also  provide systematically lower electron temperatures for these data points (\Te\,$\sim$\,30\,keV) than those derived in fits to contemporaneous flare spectra (\Te\,$\sim$\,50\,keV). 
Joint spectral fits to simultaneous \textit{Swift}/XRT, \textit{INTEGRAL}/JEMX and \textit{INTEGRAL}/ISGRI data by \citet{MKS16} obtained during the plateau observed on MJD~57194,  showed a high absorption ($N_\textrm{H}\!\approx\!1.4\times 10^{24}$\,cm$^{-2}$) over a dominant reflection component. 

Adding an   absorption  component (\textsc{tbabs}; $N_{\rm H}$ $\gtrsim$ 10$^{24}$ cm$^{-2}$) to the fits to the plateau spectra (as described in detail in Sect. 3.4), we obtain \Te\ $\sim$\,50\,keV, in better agreement with the values derived during  contemporaneous flares. 

\subsubsection{{\sc epoch 3}: Major flares and onset of outburst decay}
\label{sect:epoch3}
Between MJD 57199.05 and MJD 57200.10 we observed two major flares separated by a long X-ray plateau, similar to those seen in {\sc epoch2}.
The first flare (MJD 57199.05--57199.15; Fig. \ref{fig:zoom_hard}) reached a peak flux \Fx~$\sim\!55\!\times\!10^{-8}$\,erg\,cm$^{-2}$\,s$^{-1}$. During this flare, the spectrum was hard ($\Gamma\,\le$\,1.7), contrary to the softer flares detected in \textsc{epoch~2}. We measure roughly constant electron temperatures (\Te\,$\sim$\,50\,keV) and an optical depth $\tau$ in the range [4--5.5], which decreased  as the flare proceeded. During the subsequent X-ray plateau
a flux \Fx\ $\sim\!5\!\times\!10^{-8}$\,erg\,cm$^{-2}$\,s$^{-1}$ was measured (Fig. \ref{fig:zoom_hard}). The plateau lasted $\sim$\,0.15\,day.

The plateau was followed by a major X-ray flare (MJD 57199.50--57199.80; Fig. \ref{fig:zoom_soft}) during which the source  reached the highest fluxes measured during the 2015 outburst (\Fx$~\sim~80 \times\!10^{-8}$\,erg\,cm$^{-2}$\,s$^{-1}$). 
The flare had two peaks, separated by a $\sim$\,1\,hour  drop in flux (from $\sim$\,70\, to  $\sim$\,20\,$\times$\,10$^{-8}$\,erg\,cm$^{-2}$\,s$^{-1}$ and back to $\sim$\,80\,$\times$\,10$^{-8}$\,erg\,cm$^{-2}$\,s$^{-1}$). 
Over the flare rise and decay we find a Comptonized, soft, X-ray spectrum that softened as \Fx\ increased, and hardened as \Fx\ decreased. \Te\ also evolved in correlation with the flux variations, and reached values above $\sim$\,130\,keV ({\sc nthcomp})  or $\sim$\,90\,keV  ({\sc compps}) during the peak of the flare.  In some occasions around the peak of the flare \Te\ is  not constrained by our fits (\textsc{nthcomp}).

During the flare decay we find an abrupt drop in flux (from $\sim\!45$ to $\sim\!15$ $\times$\,10$^{-8}$\,erg\,cm$^{-2}$\,s$^{-1}$), which happened in less than half an hour. The drop in flux was accompanied by a transition to  harder spectra, characterized by a roughly constant power-law index ($\Gamma$\,$\gtrsim$\,1.5),  increasing  \Te\ and decreasing $\tau$. After the transition, the flux decay continued at a roughly constant $\Gamma$.
As the flux evolved towards quiescence values, the spectrum softened again, \Te\ was unconstrained and the optical depth, $\tau$, decreased. The two lowest flux spectra in Fig. \ref{fig:spe.example}b correspond to this period.

\begin{figure*}
\begin{tabular}{c}
\includegraphics[width=6cm,angle=90]{figures/param_distr_nthcomp_400mil.eps}\\
\includegraphics[width=6cm,angle=90]{figures/param_distr_compps_400mil.eps}\\
\end{tabular}
\caption{Distribution of the spectral parameters obtained in the fits to the IBIS/ISGRI spectra of V404 Cyg analyzed in this work. Grey bars are used to describe the total parameter distribution.
Green, blue and red symbols are used to highlight  the  best-fitting model, according to our model selection criteria (Sect. \ref{sec:spectral.modelling}). Blue: Comptonization models with  constrained \Te\ (Fig.~\ref{fig:spe.example}a) further divided into hard  ($\Gamma\,\le\,$1.7; dark blue) and soft spectra  ($\Gamma\,>$\,1.7; light blue). Green: Comptonization models with unconstrained \Te\ (Fig.~\ref{fig:spe.example}b). Red: $p$--value $<$0.05 fits.  (Fig.~\ref{fig:spe.example}c).
In the upper panels of the figure (a-c) we show the distribution of the {\sc nthcomp} parameters, while in the lower panels we display the {\sc compps} parameters.}
\label{fig:param_distr}
\end{figure*}

\subsection{Parameter distributions}
\label{sect:params.histograms}
The distributions of the spectral parameters derived in this analysis are displayed in Fig. \ref{fig:param_distr}. 
Note that the integration times of the spectra are flux dependent, and consequently the parameter distributions are skewed towards higher fluxes, when the integration times are shorter, and therefore the sampling is more frequent.

The \Fx\ distribution is shown in Fig. \ref{fig:param_distr}a and \ref{fig:param_distr}d.
We measure fluxes in the range [0.01--80] $\times$\,10$^{-8}$\,erg\,cm$^{-2}$\,s$^{-1}$ with a peak in the distribution at $\sim$\,50\,$\times$\,10$^{-8}$\,erg\,cm$^{-2}$\,s$^{-1}$. We do not find hard Comptonized spectra ($\Gamma$\,$\le$\,1.7)  when \Fx~$\gtrsim\!50\!\times\!10^{-8}$\,erg\,cm$^{-2}$\,s$^{-1}$. We find soft Comptonized spectra ($\Gamma\,>$\,1.7) for fluxes in the range [4--80] ($\times$\,10$^{-8}$\,erg\,cm$^{-2}$\,s$^{-1}$). Comptonized spectra with unconstrained electron temperatures are predominantly found  at the lowest and highest fluxes.
Spectra not compatible with Comptonized models are 
predominant in the range of fluxes [1.5--6]\,$\times$\,10$^{-8}$\,erg\,cm$^{-2}$\,s$^{-1}$, with a peak in the distribution at 5\,$\times$\,10$^{-8}$\,erg\,cm$^{-2}$\,s$^{-1}$.

The photon index ($\Gamma$; Fig. \ref{fig:param_distr}b) was derived using the  {\sc nthcomp} model. The distribution of $\Gamma$ values is  asymmetric, with a peak at $\Gamma\!\sim\!1.7$\, and a tail extending to  $\Gamma\!\sim\!3.0$.  We measure $\Gamma$  values in the range [1.5--2.4] for the Comptonized spectra with constrained \Te. All the Comptonized spectra  softer than $\Gamma$\,$\gtrsim$\,2.4  display unconstrained \Te\ in our \textsc{nthcomp} fits. 
There are also a fraction of hard spectra ($\Gamma$\,$\sim$\,1.7) with unconstrained \Te. 
The spectra detected during X-ray plateaus display $\Gamma$ values in the range [1.6--3.0], with a peak in the distribution at $\Gamma$\,=\,1.8. 

The  optical depth of the Comptonizing plasma ($\tau$; Fig. \ref{fig:param_distr}e), derived using  {\sc compps}  shows a large scatter in values, which are in the range [$0.1-5.0$]. The hard and soft spectra have different $\tau$ distributions. For the hard spectra ($\Gamma$\,$\le$\,1.7) we derive  $\tau$ values in the range [3--5.5]. For the softer spectra (1.7$<$\,$\Gamma$\,$\le$\,2.4) we derive $\tau$ values in the range [0.1--4.5].

The electron temperatures that we derive using the  {\sc  nthcomp}  and {\sc compps} models (Figs. \ref{fig:param_distr}c and \ref{fig:param_distr}f) display similar distributions, with a narrow peak at moderate temperatures ({\sc nthcomp}: 45\,keV; {\sc compps}: 35\,keV) and a tail extending up to $\sim$\,150\,keV.  However, the distribution of \Te\   derived using  {\sc nthcomp} is broader than the distribution of values obtained from the {\sc compps} fits.
We find  more spectra with unconstrained \Te\ using {\sc nthcomp}  than using {\sc compps}, probably due to the systematically lower \Te\ values derived using {\sc compps}. 
The electron temperatures derived for Comptonized hard ($\Gamma\,\le$\,1.7) and soft  spectra ($\Gamma$\,$>$\,1.7) have consistent values, but we note that the tail of the \Te\ distribution extends to  higher  energies  for the soft Comptonized spectra than for the hard ones.
The fits to the spectra obtained during X-ray plateaus provide broad \Te\ distributions. They peak at  lower energies than those derived for the Comptonized spectra ($\sim$\,25\,keV) and extend up to $\sim$\,130\,keV.

\subsection{Parameter relations}
\label{sect:params.relations}

The flux dependencies of the photon index and optical depth are presented in the  $F_\textrm{x}-\Gamma$  and $F_\textrm{x}-\tau$ diagrams  (Fig. \ref{fig:nthcomp_relat}a, \ref{fig:nthcomp_relat}d).
We find that the hard spectra in our sample ($\Gamma\!\le\!1.7$) occupy a region  in the \Fx--$\Gamma$ diagram reminiscent of  the  hard state in the BHB  Hardness-Intensity Diagram (HID; \citealt{HWvK01,B04,FBG04,DFK10}). In this region, which we will call hereafter the \textit{hard branch}, 
  the spectrum softens (from $\Gamma\approx$\,1.5 to $\Gamma\approx$\,1.7)
while the flux increases, and the optical depth increases with increasing flux from $\tau\!\sim\!3$ to $\tau\!\sim\!5.5$ (Fig. \ref{fig:nthcomp_relat}d).
The hard spectra also occupy defined regions in the \Fx--\Te, $\Gamma-$\Te, and $\tau-$\Te\ diagrams (Fig. \ref{fig:nthcomp_relat}b, \ref{fig:nthcomp_relat}e, \ref{fig:nthcomp_relat}c, \ref{fig:nthcomp_relat}f).  In these \textit{hard branch}(es), \Te\ is anti-correlated with \Fx\ (figure \ref{fig:nthcomp_relat}b,  \ref{fig:nthcomp_relat}e). We also find that \Te\ is anti-correlated with $\Gamma$ and $\tau$ (\ref{fig:nthcomp_relat}c,  \ref{fig:nthcomp_relat}f).
 
The soft ($\Gamma$\,$>$\,1.7) spectra in our sample,  detected during the brightest X-ray flares, occupy a distinct region in the $F_\textrm{x}-\Gamma$ diagram  (Fig. \ref{fig:nthcomp_relat}a) which we will call hereafter the \textit{soft flaring branch}. 
In the \textit{soft flaring branch} the spectrum still softens as the flux increases (Fig. \ref{fig:nthcomp_relat}a) even though most of the  parameter dependencies are reversed with respect to the \textit{hard branch}: $\tau$ is seen to decrease with increasing flux  (Fig. \ref{fig:nthcomp_relat}d); $F_\textrm{x}$ and \Te\  are correlated (Fig. \ref{fig:nthcomp_relat}b,  \ref{fig:nthcomp_relat}e) and the $\Gamma$--\Te\ dependency is also reversed. In the $\tau$--\Te\ diagram, we find a range of $\tau$ values ($\tau\approx$[2.5--4]) for which  \Te\ displays a roughly constant value ($\sim$\,40\,keV). Below $\tau$\,$\lesssim$\,2.5, \Te\ and $\tau$ are  anti-correlated  (Fig. \ref{fig:nthcomp_relat}f).

For spectra softer than $\Gamma$\,$\gtrsim$\,2.4 the spectral fits using \textsc{nthcomp} do not provide constrained electron temperatures. These soft spectra are detected at the highest ($\gtrsim$\,20\,$\times$\,10$^{-8}$\,erg\,cm$^{-2}$\,s$^{-1}$) and lowest ($\lesssim$\,2\,$\times$\,10$^{-8}$\,erg\,cm$^{-2}$\,s$^{-1}$)  \Fx\ values. When detected at the highest fluxes, they occupy regions in the \Fx$-\Gamma$ diagram (Fig.~\ref{fig:nthcomp_relat}a) reminiscent of the HID  \textit{ultra-luminous state} \citep{MBS14}.

Finally, we also observe that the spectra detected during X-ray plateaus occupy separate regions in all these diagrams, distinct from the Comptonized branches, which confirms our classification of these spectra in a separate category. We  will call these regions \textit{plateau branch(es)}.

\begin{figure*}
\begin{tabular}{l}
\includegraphics[width=5.8cm,angle=90]{figures/param_relations_nthcomp_400mil.eps}\\
\includegraphics[width=5.8cm,angle=90]{figures/param_relations_compps_400mil.eps}
 \end{tabular}
 \caption{Relations between the parameters derived from our spectral fits, using the {\sc nthcomp} (panels a--c) and the {\sc compps} models (panels d--f). Green, blue and red symbols are used to highlight  the  best-fitting model, according to our model selection criteria (Sect. \ref{sec:spectral.modelling}). Blue: Comptonization models with  constrained \Te\ (Fig.~\ref{fig:spe.example}a) further divided into hard  ($\Gamma\,\le$\,1.7; dark blue) and soft spectra  ($\Gamma$\,$>$\,1.7; light blue). Green: Comptonization models with unconstrained \Te\ (Fig.~\ref{fig:spe.example}b). Red: $p$--value $<$0.05 fits.  (Fig.~\ref{fig:spe.example}c).
{\bf Panel a)} $F_\textrm{x}-\Gamma$ diagram. We find a \textit{hard branch} (dark blue points) where $F_\textrm{x}$ and $\Gamma$ are correlated, 
 similar to the hard state in the BHB HID.  We also find a \textit{soft flaring branch} (light blue points), where the source exhibits the highest fluxes in the outburst, and the $F_\textrm{x}$ and $\Gamma$ correlation persists. We identify the \textit{soft flaring branch} with the BHB \textit{intermediate states}.
 We  also identify the softest spectra in our sample ($\Gamma\gtrsim\! 2.4$) with unconstrained \Te\  with a  tentative \textit{ultra-luminous} state (green points).
{\bf Panel d)} $F_\textrm{x}-\tau$ diagram. In the \textit{hard branch} $\tau$ and  $F_\textrm{x}$ are correlated, while in the  \textit{soft flaring branch} the relation reverses and these parameters  are anti-correlated.
{\bf Panel b, e)} \Fx--\Te\ diagrams ({\sc nthcomp} and {\sc compps}). In the \textit{hard branch}
 \Fx\ and \Te\ are anti-correlated, while in the \textit{soft flaring branch}
\Fx\ and \Te\ are correlated. 
{\bf Panel c)} $\Gamma-$\Te\ diagram.
In the \textit{hard branch} $\Gamma$ and \Te\ are anti-correlated: as the spectrum softens, the electrons cool down. In the \textit{soft flaring branch} $\Gamma$ and \Te\ are correlated: as the spectrum softens, we derive higher electron temperatures,  \Te\ becomes eventually unconstrained for $\Gamma >2.4$.
{\bf Panel f)} $\tau-$\Te\ diagram.
In the \textit{hard branch} $\tau$ and \Te\ are anti-correlated. 
In the \textit{soft flaring branch} we observe a moderate anti-correlation between $\tau$ and \Te\ for $\tau$ in the range [2.5--5] and an anti-correlation between $\tau$ and \Te\ for $\tau<2$, until \Te\ becomes eventually unconstrained as $\tau$ approaches zero.}
 \label{fig:nthcomp_relat}
\end{figure*}
 
\begin{figure*}
\includegraphics[width=12cm,angle=90]{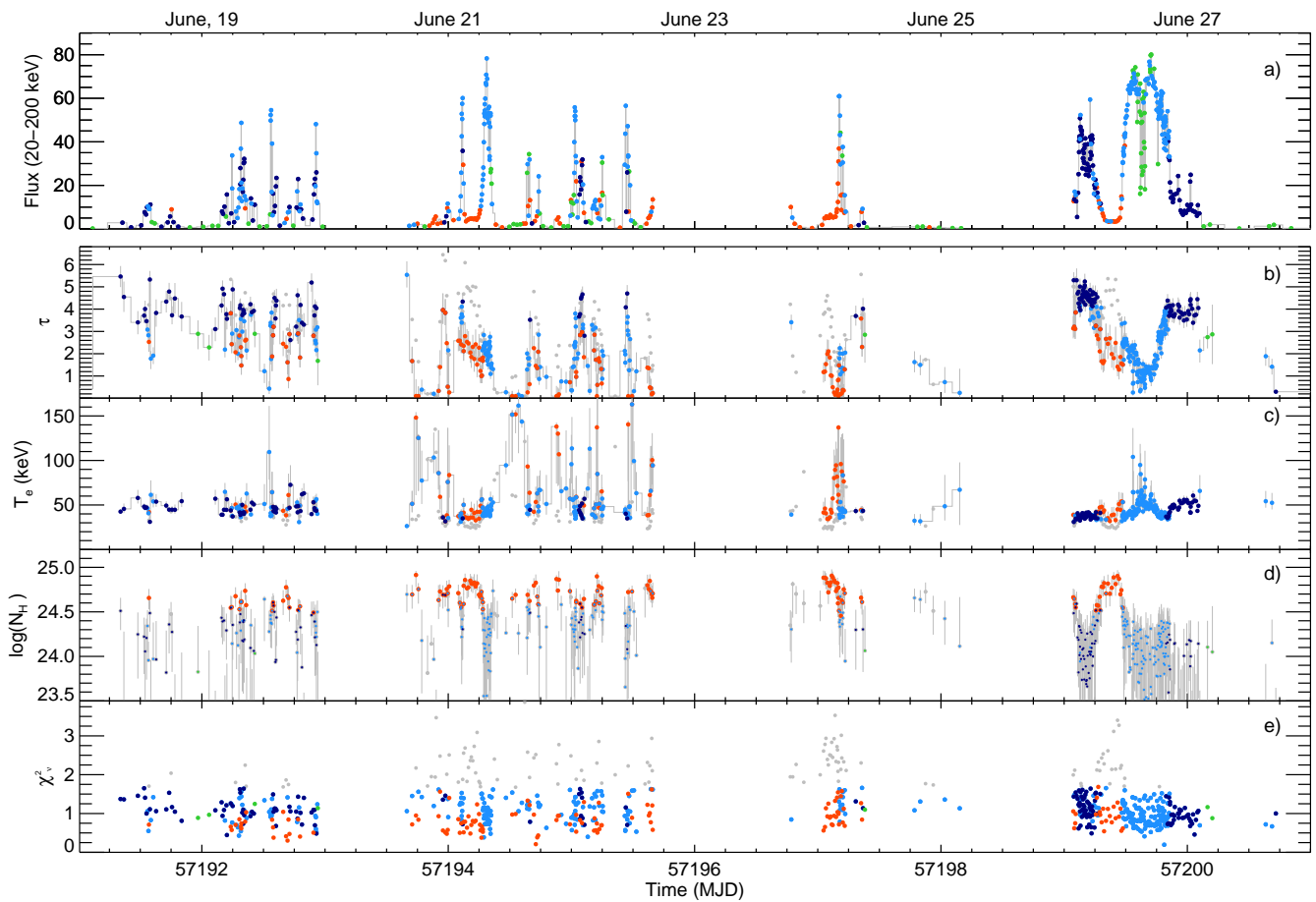}
\caption{Time evolution of the flux and spectral parameters of V404~Cyg during the June 2015 flaring episodes. Green and blue symbols are unchanged with respect to Fig. \ref{fig:overall_lcr}. Red symbols highlight the parameters obtained fitting the data with a Comptonized model (\textsc{compps}) modified by variable absorption. 
The BIC model selection (see text) favors this model only for the spectra detected during the X-ray plateaus.  
For comparison, we also show the parameters obtained in the fits to these spectra fixing the absorption to interstellar values (grey symbols in panels b, c and e). 
\textbf{Panel a)} source flux (20--200 keV) in units of 10$^{-8}$\,erg\,cm$^{-2}$\,s$^{-1}$. \textbf{Panel b)} optical depth, $\tau$, \textbf{Panel c)} electron temperature, \Te. \textbf{Panel d)} $N_\textrm{H}$ values obtained for the whole data set when leaving $N_\textrm{H}$ as a free parameter. Note that only when $N_\textrm{H}\gtrsim\,10^{24}$ cm$^\textrm{-2}$ it modifies substantially the spectral shape in the IBIS/ISGRI energy range, and can be properly constrained in our spectral fits. This happens during the X-ray plateaus. \textbf{Panel e)} $\chi^{2}_{\nu}$ value for every fit.}
\label{fig:overall_lcr2}
\end{figure*}

\begin{figure*}
\includegraphics[width=14.5cm,angle=90]{figures/Model_comparison_only_refl.eps}
\caption{Same as Fig. \ref{fig:overall_lcr2}, but for a Comptonized model (\textsc{compps}) modified by variable absorption and reflection.
\textbf{Panel a)} source flux (20--200 keV) in units of 10$^{-8}$\,erg\,cm$^{-2}$\,s$^{-1}$. \textbf{Panel b)} optical depth, $\tau$, \textbf{Panel c)}  electron temperature, \Te. \textbf{Panel d)} $N_\textrm{H}$ values obtained for the whole data set when leaving $N_\textrm{H}$ as a free parameter.  \textbf{Panel e)} Reflection fraction (Parameter R in \textsc{compps}). \textbf{Panel f)} $\chi^{2}_{\nu}$ value for each fit.}
\label{fig:overall_lcr3}
\end{figure*}

\subsection{Fits to the plateau spectra using variable absorption}
\label{Var_abs}

To study the poor fitting spectra detected during X-ray plateaus, we also considered the possibility that these were the result of obscuration of the primary X-ray emitting region. 
Absorption by Compton-thick material ($N_{\rm H} \gtrsim\,10^{24}$ cm$^\textrm{-2}$) can substantially reduce the source flux for energies up to $\sim$\,30--40\,keV, and also change the spectral slope and produce a global flux reduction by a factor 10 in the 40--300\,keV energy range \citep{MY09}. Joint spectral fits to simultaneous \textit{Swift}/XRT and \textit{INTEGRAL}/JEMX+ISGRI spectra during the X-ray plateau around MJD 57195  by \citet{MKS16} already showed that the broad band spectrum of V404~Cyg  around that period was compatible with heavily absorbed Comptonized emission  \mbox{($N_\textrm{H}\sim~1-3\times10^{24}\,\textrm{cm}^{-2}$)} with a prominent scattered component.

For the above reasons, we performed two additional fitting runs using {\sc compps}. In the first fitting run, we left  $N_{\rm H}$ as a free parameter. The results of these fits are presented in Fig.~\ref{fig:overall_lcr2}. In the second fitting run we also allowed the Compton reflection amplitude parameter in \textsc{compps} to vary freely. 
The results obtained in these fits are presented in Fig.~\ref{fig:overall_lcr3}.
In both cases, we compared the results obtained in these fits against those obtained using a model where the absorption was fixed to interstellar values, as described in the previous sections. We computed the BIC for each fit, and again applied the $\Delta$BIC$>$6 criterium to select the model better describing the data.
Also, the $p$--value of the fit with respect to the data was verified in these new fits.\\
\indent
For the spectra obtained during the X-ray plateaus, the $\Delta$BIC model selection criterium favors a Comptonized model with $N_\textrm{H}$ values in excess of the interstellar one ($N_\textrm{H}\,\gtrsim\,5\times\,10^{24}$\,cm$^{-2}$, see Figs. \ref{fig:overall_lcr2}, \ref{fig:overall_lcr3}) regardless whether reflection is considered or not. However, $N_\textrm{H}$ is constrained better when no reflection is considered (see Fig. \ref{fig:overall_lcr2}).
The validity of the models is supported by the $p$--value of the fit with respect to the data, which is now  $p$--value~$\ge\,0.05$ for the fits to the plateau spectra, when the absorption is left to vary freely, with or without reflection.
During X-ray flares, the BIC model selection favours a Comptonized model where the absorption is fixed to the interstellar values. The energy range analyzed in this work (20--200\,keV), only allows 
firm constraints on the most extreme $N_\textrm{H}$ values ($N_\textrm{H}\,\gtrsim\,5\times\,10^{24}$\,cm$^{-2}$). Although we observe an evolution of the derived $N_\textrm{H}$ values in anti-correlation with the flux evolution (i.e. $N_\textrm{H}$ reaches the highest values during the X-ray plateaus and decreases during the X-ray flares).\\
\indent
Using solely the 20--200\,keV energy range of our spectra, we cannot constrain simultaneously  $N_\textrm{H}$ and the reflection contribution to the source spectra (quantified using the $R$ parameter in \textsc{compps}, Fig. \ref{fig:overall_lcr3}e), as shown by the large uncertainties in the determination of both parameters (Fig. \ref{fig:overall_lcr3}d, e). 
The BIC model selection criterium favours a Comptonized model with variable column density over a Comptonized model with variable column density and variable reflection. Only extending the fitting range to lower energies we will be able to model simultaneously the absorption and reflection parameters in our fits.\\
\indent
Finally, we note that with the spectral fits presented in this section, the peculiar increase of $\tau$, and simultaneous decrease of \Te\ derived during the X-ray plateaus largely disappears, (see Figs. \ref{fig:overall_lcr},  \ref{fig:overall_lcr2}, \ref{fig:overall_lcr3})  suggesting that the systematic changes of these parameters were the result of an inaccurate modeling.

\section{Discussion}
The light curve of V404~Cyg during the June 2015 outburst does not display the typical features of the standard BHB light curves (e.~g.  \citealt{CSL97,RMcC06}).
Similarly, the soft X-ray spectra of V404~Cyg is remarkably different from the spectra of other BHBs, mostly due to extreme intrinsic absorption \citep{MKS16} also seen in the 1989 outburst \citep{ZDS99}.
However, when we look at the source spectra in hard X-rays (above 20~keV), we find some similarities between V404~Cyg and other BHBs. 

\subsection{Hard branch}
We have identified a \textit{hard branch} in the F$_X$--$\Gamma$ diagram of V404~Cyg, reminiscent of the \textit{hard state} branch of the BHB  HID \citep{HWvK01,B04,FBG04,DFK10,BM16}. 
The \textit{hard branch} is occupied by the hardest spectra in our sample ($\Gamma\!\le\!1.7$). 
In the \textit{hard branch}, the spectrum of V404~Cyg gradually softens and \Te\ decreases from \Te$\sim 80$~keV or unconstrained (\textsc{nthcomp}) down to about \Te$\sim 40$~keV as the flux increases (Fig. \ref{fig:nthcomp_relat}a, \ref{fig:nthcomp_relat}d, \ref{fig:nthcomp_relat}b, \ref{fig:nthcomp_relat}e). This \Fx--\Te\ anti-correlation was observed by \citet{NFB15} \citet{RJB15} in the analysis of IBIS/ISGRI and SPI data obtained during rev. 1554 (\textsc{epoch 1}) and by \cite{JWHH16} using \textit{FERMI}/GBM data. Similar anti-correlations are also found in other BHBs in the \textit{hard state} \citep{ENC98,WZG02,RCT03,MBH09,KVT16}, supporting our identification of the \textit{hard branch} with the \textit{hard state} of prototypical BHBs \citep{BM16}.
The \Fx--\Te\  anti-correlation in the \textit{hard branch} can be explained by 
the truncated disk/hot inner flow model \citep{EMcCN97,MP07}, which assumes that at low luminosities the accretion disk is truncated at distances between a few tens and a few thousand gravitational radii from the BH, and only a small fraction of disk photons reach the hot flow/comptonizing medium. The X-ray spectrum would then be produced by pure synchrotron self-Compton emission (SSC) in an hybrid (thermal plus non-thermal) Comptonizing medium \citep{PV09,MB09,VVP11}. V404~Cyg may be in this regime at fluxes below $\sim\!10^{-8}$\,erg\,cm$^{-2}$\,s$^{-1}$, where we measure \Te\ in the range 60--80\,keV (or unconstrained). 
As the accretion rate increases, the  inner radius of the accretion disk moves inwards, closer to the BH, and a growing number of soft seed photons from the accretion disk enter the Comptonizing medium, gradually cooling down the population of thermal electrons \citep{PV09,MB09,VVP11}. 
Electron cooling results in softer Comptonized spectra \citep{DGK07}. 
The electron cooling could cause the observed \Fx--\Te\ anti-correlation and gradual spectral softening in the \textit{hard branch}. 
Observations of GX~339--4 \citep{WZG02} and GRO~J1655--40 \citep{JKS08} in the \textit{hard state} result in \Te\ values comparable to our measurements. However, these systems displayed lower $\tau$ values ($\tau\!\approx$2.5) than those measured for V404~Cyg ($\tau\approx$[3--5.5]) which suggests that we are dealing with an optically thicker Comptonizing medium. The  $\tau$~values we derive are comparable to those found during the 1989 outburst of V404~Cyg ($\tau\approx\!6$; \citealt{ZDS99}).
As the optical depth is expected to scale linearly with the mass accretion rate \citep{RC00}, the presence of an optically thick Comptonizing medium may be connected to V404 Cyg emitting closer to the Eddington limit than other BHBs in the \textit{hard state}.

\subsection{Soft flaring branch}
In the  \textit{soft flaring branch}  the spectrum still softens as the flux increases, but most of the  parameter dependencies are reversed with respect to the \textit{hard branch}, suggesting a change in the hard X-ray production mechanism: as \Fx\ increases, the optical depth of the Comptonizining medium decreases and the spectrum softens (Fig. \ref{fig:nthcomp_relat}c), while the electron temperature increases (Fig. \ref{fig:nthcomp_relat}b,  \ref{fig:nthcomp_relat}e). These parameter relations  are similar to those observed during hard to  soft  state transitions in other BHBs \citep{PJL96, ENC98,JKS08,MBH09,dSMB13}, suggesting that the \textit{soft flaring branch} may correspond to the BHB \textit{intermediate state}. The decrease in $\tau$ observed during \textit{hard} to \textit{soft} state transitions is consistent with the material in the Comptonizing medium  condensing into a disk, or being ejected into an outflow  \citep{M16}. The detection of a  cut-off in the spectrum is indicative of a significant fraction of thermal electrons in the Comptonizing medium. \Te\ increases  as the flux progressively increases, suggesting that the injection of external (disk) photons, which cooled down the electron cloud in the \textit{hard branch} might have  ceased or its cooling effect on the electron cloud is negligible. 

The steepest spectra in our sample with $\Gamma \gtrsim 2.4$ (or $\tau \lesssim 1.0$) are the ones where \Te\ is not constrained  (\textsc{nthcomp}). Fits to these data using \textsc{compps} provide extremely low $\tau$  and high \Te\ values.
When the optical depth is low, the particles do not have time to  
thermalize between re-accelerations, and the electron distribution resembles the power-law injection function that could originate from magnetic re-connection or shock acceleration \citep{VVP11,dSMB13}.
These spectra are  detected at epochs bracketed by the largest measured \Te\ values ($\gtrsim 80$ keV), so it is likely that they are caused by the electron population becoming simultaneously hotter and/or progressively less thermal. When detected at the highest fluxes ($\gtrsim$\,20\,$\times$\,10$^{-8}$\,erg\,cm$^{-2}$\,s$^{-1}$), the  $\Gamma \gtrsim 2.4$  spectra occupy a region in the \Fx$-\Gamma$ diagram (Fig. \ref{fig:nthcomp_relat}a) reminiscent of the HID \textsl{ultra-luminous state}  \citep{DFK10}, which suggests that these  spectra could be analogous to the \textsl{ultra-luminous state} of GX~339-4  \citep{MKK91,KD16}, GS~1124--68 \citep{MIK93},  
XTE~J1550--564 \citep{SMR99b,KM04,HAD16}, GRO~J1655--40 \citep{SMR99a,JKS08}, or 4U 1630--47 \citep{AFK05} where the spectrum is a composite of a strong disc and a steep prominent Comptonized tail, with no cutoff at high energies, which may extend up to $\sim$\,1\,MeV \citep{KD16}.

When detected at the lowest fluxes, these $\Gamma \gtrsim 2.4$ spectra, could be analogous to the \textit{soft state} spectra of some BHB where faint non-thermal hard X-ray tails are detected \citep{ZMC16}. 

\subsection{X-ray plateaus}

We have  observed X-ray plateaus characterized by roughly constant fluxes  (\Fx\ $\sim\!5 \times 10^{-8}$\,erg\,cm$^{-2}$\,s$^{-1}$)  during periods lasting several hours. Often the plateaus are observed in between two or several flares (see Figs. \ref{fig:overall_lcr}, \ref{fig:zoom_plateau}).  Furthermore, we only detect X-ray plateaus in epoch 2 and 3, when the source is predominantly in the \textit{soft flaring branch}, where the spectrum is softer ($\Gamma\,>\,1.7$). We have observed that the statistically worse fits ($\chi^2_{\nu}>$1.8; see Fig. \ref{fig:overall_lcr}) tend to happen during these X-ray plateaus if we fix $N_\textrm{H}$ to the interstellar value in the direction of the source. The statistics are improved when allowing $N_\textrm{H}$ to vary in our fits ($\chi^2_{\nu}\sim$1; see Figs. \ref{fig:overall_lcr2}, \ref{fig:overall_lcr3}). In this case we derive $N_\textrm{H}$  values ($N_\textrm{H}\,\gtrsim\,5\times\,10^{24}$\,cm$^{-2}$) two orders of magnitude in excess of the interstellar $N_\textrm{H}$ value  in the direction of the source ($N_\textrm{H}\,\sim\,0.8\times\,10^{22}$\,cm$^{-2}$), which suggests that intrinsic absorption by Compton-thick material distorts the source spectrum, and results in 
the spectral shape observed during the X-ray plateaus (see Fig.~\ref{fig:spe.example}c). Due to the limited energy range available for this analysis, our fits cannot constrain simultaneously $N_\textrm{H}$ and the fraction of X-ray photons reprocessed by the Compton thick material contributing to the source spectra (i.e. the reflection fraction in \textsc{compps}). We note also that if $N_\textrm{H}$ is allowed to vary in our fits, the  spectral parameters \Te\ and $\tau$ display values consistent with measurements during nearly contemporaneous flares.

In previous studies, \citet{NFB15}, \citet{RJB15} and \citet{JWHH16} inferred very high seed photon temperatures $\sim$\,6--7\,keV, which they attributed to the jet. 
Our results show that the source spectra  can also be modeled using lower seed photons temperatures (consistent with Swift/XRT results; \citealt{MKS16}) and strong local absorption.

\subsection{Flaring activity}
Although $N_\textrm{H}$ cannot be properly constrained during the X-ray flares, when lower $N_\textrm{H}$ values are derived, it is seen to vary in anti-correlation with the evolving X-ray flux  over the whole data set analyzed here (Fig. \ref{fig:overall_lcr2}c). The fact that we derive $N_\textrm{H}\sim\,5\times\,10^{24}$\,cm$^{-2}$ only during the X-ray plateaus, and clearly lower $N_\textrm{H}$ values in the adjacent flares, suggests that the dramatic intensity drops observed during these plateaus may be partially caused by obscuration of the central source. The obscuring material can be some outflowing material, the outer regions of a Compton thick accretion disk, or a combination of both. 
Thus, the apparent flaring activity may be actually the result of a clumpy Compton-thick obscuring material becoming occasionally Compton-thin and allowing the source photons to reach the observer. The fast flare rise and decay times ($\approx$30\,min, see Fig. \ref{fig:zoom_plateau}) may actually be related to  varying partial obscuration of the central source, as previously suggested by \citet{ZDS99}, who measured $N_{\rm H}$ variability on timescales of minutes during the 1989 outburst.
Perhaps also for the same reason we find a lot of scatter in the correlations between \Fx\ and the source spectral parameters ($\Gamma,\tau$, \Te), while there is much less scatter in the various parameter correlations ($\Gamma$--\Te, $\tau$--\Te\; see Fig. \ref{fig:nthcomp_relat}). 

\section{Summary of results and conclusions}

We have fit the 20--200\,keV  IBIS/ISGRI spectra of V404~Cyg  during the June 2015 outburst using two thermal Comptonization models  (\textsc{nthcomp} and \textsc{compps}).
For the first time we have continuously measured the evolution of the properties of the Comptonizing medium during an outburst rise and decay.  We find that the system evolves through the same $\Gamma$--\Te, $\tau$--\Te\ and \Fx--\Te\ paths when the outburst rises or decays.  We have identified two clear spectral branches in the \Fx\--$\Gamma$ diagram which display characteristic parameter relations: a \textit{hard branch} and a \textit{soft flaring branch}. 

In the \textit{hard branch}, V404~Cyg shows a  hard  ($\Gamma\!\le$\!1.7) thermal Comptonized spectrum, which slowly softens as the flux increases. In the \textit{hard branch}, $\tau$ is correlated with \Fx\, while \Te\ is anti-correlated with \Fx\ and $\Gamma$. Similar parameter correlations have been observed in other BHBs in the \textit{hard state}, suggesting that the \textit{hard branch} could correspond to the HID \textit{hard state}. The observed parameter evolution can be explained in terms of  thermal Comptonization of soft seed photons by a hot electron cloud in the vicinity of the BH. The \Fx--\Te\ anti-correlation could result from the electron population  progressively cooling down as the accretion disk moves closer to the BH and more disk photons enter the Comptonizing medium. 

In the \textit{soft flaring branch} V404~Cyg shows a soft, thermal Comptonized spectrum, ($\Gamma\!>$\!1.7), which softens as the flux increases. In the \textit{soft flaring branch} \Te\ and \Fx\ are correlated, while \Fx\ and $\tau$ are anti-correlated. The parameter correlations are consistent with those observed during \textit{hard} to \textit{soft} state transitions in other sources, suggesting that these data could correspond to the \textit{intermediate state} or occasionally the \textit{ultra-luminous state}. The observed \Te--\Fx\ correlation is compatible with the predictions of SSC-models.

We have also found \textit{a plateau branch} where Comptonization models fail to describe the source spectra if $N_\textrm{H}$ is fixed to the interstellar values in the direction of the source. The fits to these spectra improve when we leave the absorption to vary freely. In this case we derive $N_\textrm{H}$  values ($N_\textrm{H}\,\gtrsim\,5\times\,10^{24}$\,cm$^{-2}$) which suggests that intrinsic absorption by Compton-thick material results in 
the spectral shape observed during the X-ray plateaus.
The obscuring material can be some outflowing material, a clumpy Compton thick accretion disk, or a combination of both. 
Thus,  we propose that the observed dramatic flaring activity seen at hard X-rays may not only be due to intrinsic source variability, but can  partly result from obscuration of the central source by Compton thick material.
The system inclination of 67$^{\circ}$ may be a key parameter in he observation of such phenomenology, not observed in other sources seen at lower or higher inclination angles.

\begin{acknowledgements}
The authors would like to thank the anonymous referee for useful comments  that  contributed  to  improve  the  paper. JJEK was supported by Academy of Finland grants 268740 and 295114, and the ESA research fellowship programme.
SEM acknowledges support from the Faculty of the European Space Astronomy Centre (ESAC).
\end{acknowledgements}
\bibliographystyle{aa}
\bibliography{v404cyg.bib}

\begin{appendix}
\section{Closer view to some flares}

\begin{figure*}
\includegraphics[width=17.0cm,angle=90]{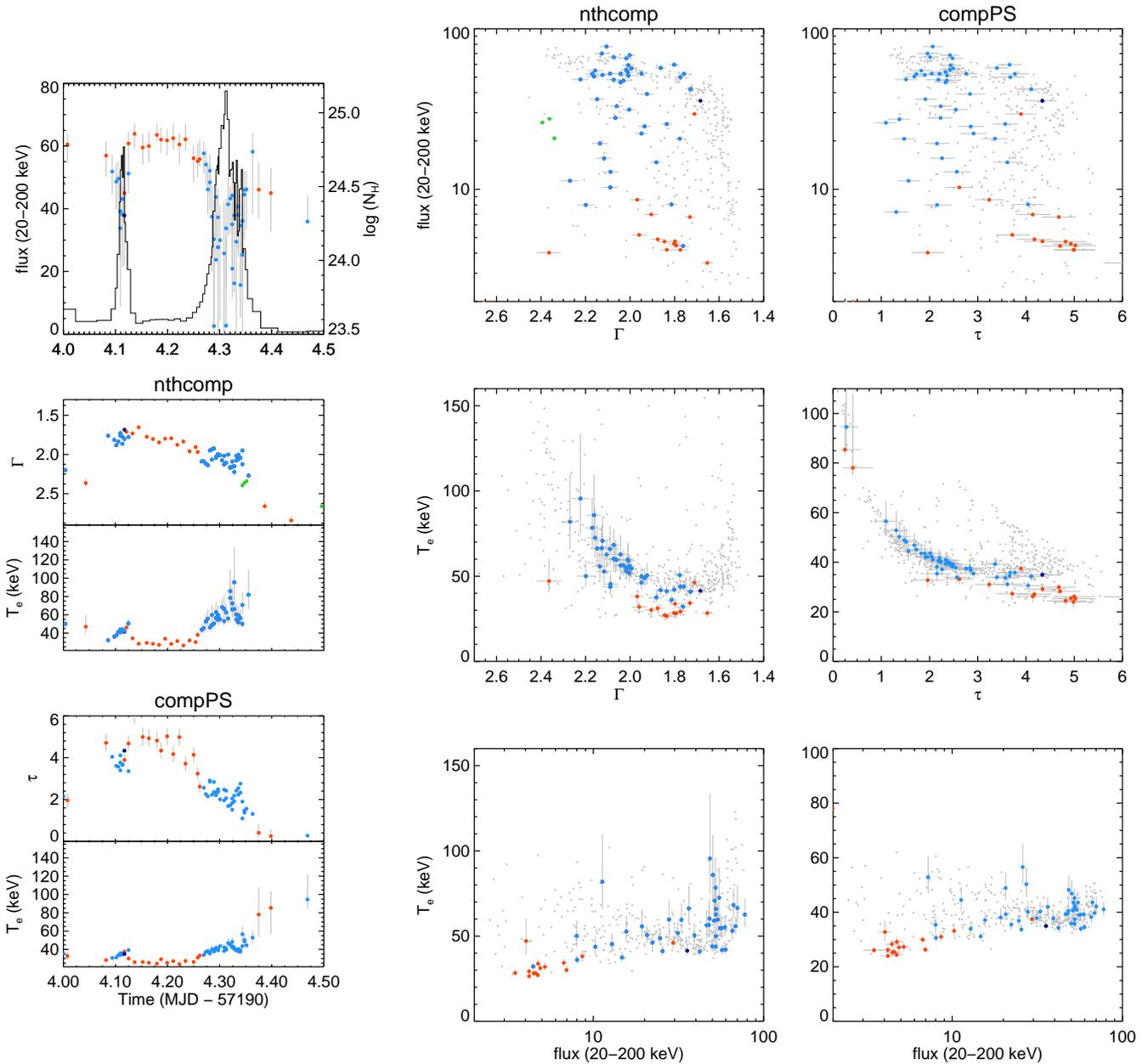}
\caption{Close view to the properties of the X-ray plateau detected on MJD 57194. In the upper left panel we provide the evolution of the source flux, \Fx\ in the 20--200\,keV energy range, ($\times10^{-8}$\,erg\,cm$^{-2}$\,s$^{-1}$), together with the evolution of the absorption column, $N_\textrm{H}$, as derived leaving it as a free parameter in our spectral fits (see Sect.~\ref{Var_abs}). The plateau happens in the interval MJD 57194.0--57194.4. It is characterised by a roughly constant flux  \Fx
\,$\sim\!5\!\times\!10^{-8}$\,erg\,cm$^{-2}$\,s$^{-1}$  and is interrupted by two major flares around MJD 57194.1 and 57194.3 with peak fluxes of \Fx
\,$\sim\!60$ and $80\!\times\!10^{-8}$\,erg\,cm$^{-2}$\,s$^{-1}$.  We observe roughly constant $N_\textrm{H}$ values over the X-ray plateau ($N_\textrm{H}\,\sim\,5\,\times\,10^{24}\,$cm$^{-2}$). Systematically lower $N_\textrm{H}$ values are measured during the two X-ray flares. However we note here that $N_\textrm{H}$ cannot be properly constrained by our fits to the flare spectra. The additional plots in the left side column  provide the time evolution of spectral parameters during this time interval, derived in this work using Comptonization models (\textsc{nthcomp} and \textsc{compps}) and absorption fixed to the interstellar values. The plots in the middle and right columns describe the relations between these spectral parameters derived using  \textsc{nthcomp} (middle column) and \textsc{compps}  (right column).
The $\Gamma$ and $\tau$ parameters derived during the plateau display values consistent with the trend observed during the flares, while the electron temperatures have systematically lower values than measured in the flare spectra.
The parameters derived during the peaks of the flares occupy the \textit{soft flaring branch}, and display typical spectral characteristics of this region (see Sect. \ref{sect:params.relations}), while the flare rise and decays, still very likely affected by obscuration, occupy intermediate regions between the \textit{soft flaring branch} and the \textit{plateau branch} in the $\Gamma$--\Te\,  $\tau$--\Te\ a \Fx--\Te\ diagrams.}

\label{fig:zoom_plateau}
\end{figure*}

\begin{figure*}
\includegraphics[width=17.0cm,angle=90]{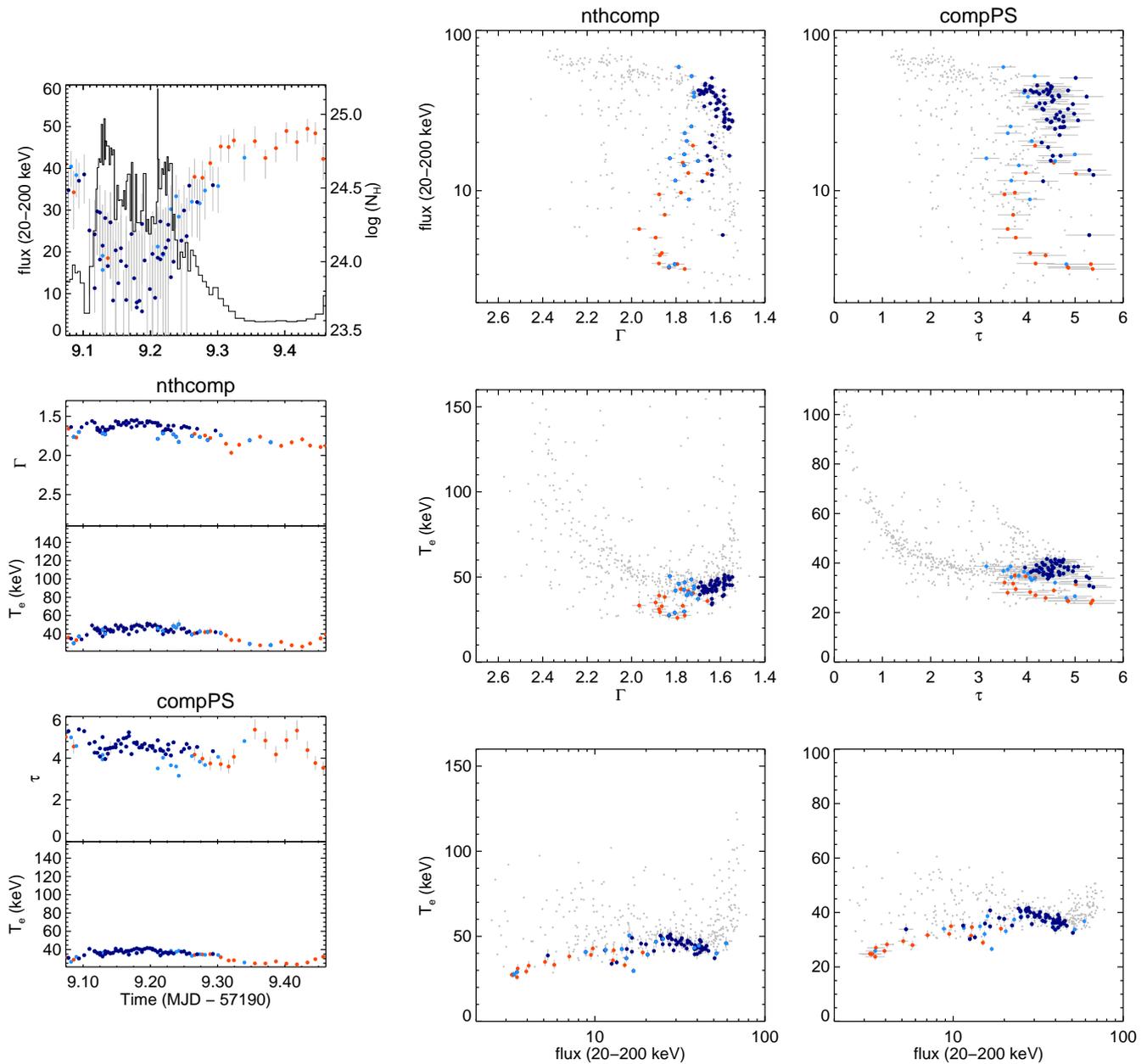}
\caption{Close view to the properties of the hard X-ray flare detected around MJD 55199. The flare is followed by an X-ray plateau also shown in the figure. In the upper left panel we provide the evolution of the source flux, \Fx\ in the 20--200\,keV energy range, ($\times10^{-8}$\,erg\,cm$^{-2}$\,s$^{-1}$), together with the evolution of the absorption column, $N_\textrm{H}$, derived when leaving it as a free parameter in our spectral fits.  
The additional plots in the left column provide the time evolution of the system flux and spectral parameters during this time interval, as derived using Comptonization models (\textsc{nthcomp} and \textsc{compps}) fixing the absorption to the interstellar values. 
The plots in the center and right side of the figure describe the relations between these spectral parameters derived using  \textsc{nthcomp} (middle panel) and \textsc{compps}  (right panel). The   dramatic changes in flux measured during the X-ray flare are accompanied by little variations in $\Gamma$ and $\tau$. The data points corresponding to this flare  occupy the brightest \Fx\ regions of the \textit{hard branch} (see Sect. \ref{sect:params.relations}) in the \Fx\--$\Gamma$, \Fx\--$\tau$ and \Fx\--\Te\ diagrams. In these diagrams  \Te\ is anti-correlated with \Fx\  and $\Gamma$. Some, brief transitions to the \textit{soft flaring branch} are observed during the peak of the flare. During the peak of the flare, $N_\textrm{H}$ cannot be properly constrained by our spectral fits.
Following the peak, \Fx\ gradually drops to an X-ray plateau, the spectrum gradually softens, \Te\ decreases and $\tau$ displays high variability. As the flux decays we observe a gradual increase in $N_\textrm{H}$, which reaches values close to 10$^{25}$\,cm$^{-2}$ when the system reaches the bottom of the X-ray plateau.
 The various parameter correlations reverse in the \textit{plateau branch} with respect to the trends observed in the \textit{hard branch}.}
\label{fig:zoom_hard}
\end{figure*}

\begin{figure*}
\includegraphics[width=17.0cm,angle=90]{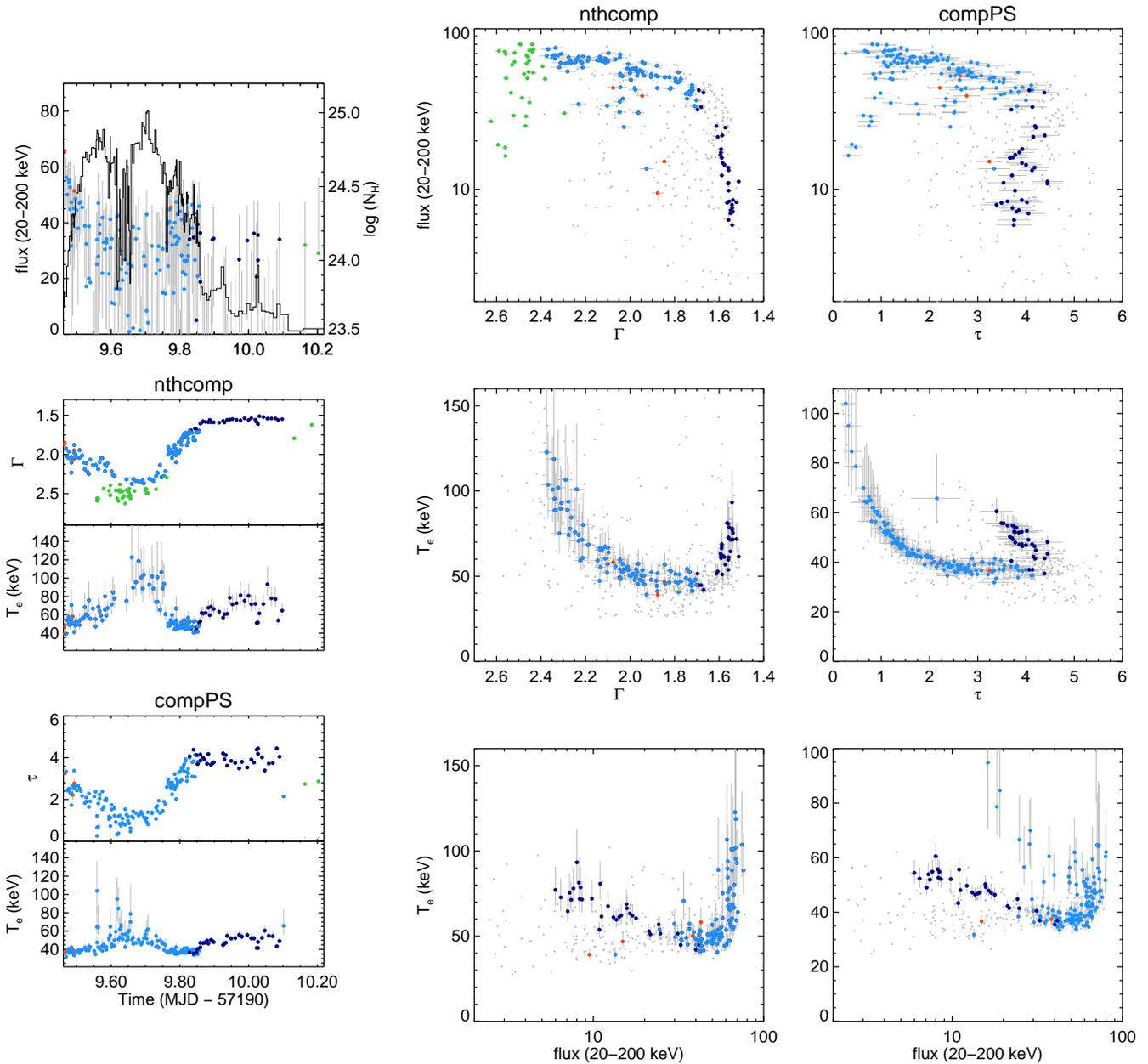}
\caption{Close view to the properties of the major X-ray flare detected around MJD 57200.
In the upper left panel we provide the evolution of the source flux, \Fx\ in the 20--200\,keV energy range ($\times10^{-8}$\,erg\,cm$^{-2}$\,s$^{-1}$), together with the evolution of the absorption column density, $N_\textrm{H}$, derived when leaving it as a free parameter in our spectral fits. Note that in this bright flare $N_\textrm{H}$ cannot be properly constrained using only the IBIS/ISGRI energy range. The additional plots in the left side column  provide the time evolution of the system flux and spectral parameters derived using Comptonized models (\textsc{nthcomp} and \textsc{compps}) fixing $N_\textrm{H}$ to the interstellar values in the direction of the source. The plots in the middle and right columns describe the relations between the different spectral parameters measured during this time interval using  \textsc{nthcomp} (middle columns) and \textsc{compps}  (right columns). During rise and decay, the flare displays parameter correlations characteristic of the \textit{soft flaring branch}: $\Gamma$ and  \Te\ are correlated with \Fx\ while $\tau$  is anti-correlated with \Fx. Around the  peak of the flare, the spectra soften above $\Gamma\!\sim\!2.4$ and \Te\ cannot be constrained using \textsc{nthcomp}. Spectral fits to these data using \textsc{compps} provide (constrained) \Te\ values (\Te\ $\gtrsim\!100$\,keV), and the lowest  $\tau$ values found in this work ($\tau\!\lesssim\!1$). These spectra occupy a region in the $\Gamma$--\Fx\ diagram reminiscent of the \textit{ultra-luminous state} in the BHB HID. These spectra also occupy a separate branch in the \textsc{compps} \Fx--\Te\ diagram. After the peak of the flare, as \Fx\ decreased and the spectrum hardened below  $\Gamma\!\sim\!2.4$ the system returned to the \textit{soft flaring branch}, and \Fx\ started to decrease from the peak values.
Around MJD~57199.85, when the system entered the \textit{hard branch},  a dramatic drop in flux was observed  (from \Fx $\sim\!50$ to \Fx $\sim\!15\!\times\!10^{-8}$\,erg\,cm$^{-2}$\,s$^{-1}$ in about half an hour), as the system entered the \textit{hard branch}. After this transition, the system started the decay to quiescence, displaying the characteristic \textit{hard branch} correlations observed during the flare rise, but for a decreasing flux: the optical depth increased and the spectrum gradually hardened as the flux decayed, while the electrons were seen to gradually heat up.}
\label{fig:zoom_soft}
\end{figure*}

\end{appendix}

\end{document}